\documentclass[amsmath,amssymb,showpacs,draft,preprint,floatfix]{revtex4}

\usepackage{graphicx}
\usepackage{latexsym}
\begin{document}
\title{Coherent states for time dependent harmonic oscillator: the step function}

\author{H. Moya-Cessa and M. Fern\'andez Guasti}
\affiliation{INAOE, Coordinaci\'on de Optica, Apdo. Postal 51 y
216, 72000 Puebla, Pue., Mexico}
\date{\today}

\begin{abstract}
We study the time evolution for the  quantum harmonic oscillator
subjected to a sudden change of frequency. It is based on an
analytic solution to the time dependent Ermakov equation for an
approximated step function, and therefore, ours is a continuous
treatment that differs from former studies that have limiting
procedures for times when the step occurs.
\end{abstract}

 \maketitle
\section{Introduction}
The problem of the harmonic oscillator with time dependent
frequency has received considerable attention over the years
\cite{several}. In particular Kiss {\it et al.} \cite{kiss} have
studied the problem of the time dependent frequency given by a
step function (see also \cite{agar}). They have solved the problem
by dividing it in two regions for the two different frequencies of
the step function and matching later the solutions for the two
regions. Here we would like to study the same problem but using a
continuous approach, by using instead an invariant formalism. This
will allow us to obtain analytic solutions that show squeezing
(departing from an initial coherent state) or not depending on
time. We will also show that, in a transformed Hilbert space, an
initial coherent state remains coherent during the evolution.

\section{Coherent states for constant frequency}
Let us first consider the Hamiltonian for the harmonic oscillator
with unitary mass, $m=1$,  and unitary frequency, $\omega_0=1$

\begin{equation}
\hat{H} =\frac{1}{2}(\hat{q}^2+\hat{p}^2).
\label{unity}%
\end{equation}
We can define annihilation and creation operators as
\begin{equation}
\hat{b} =\frac{1}{\sqrt{2}}( \hat{q}+i\hat{p} ), \qquad
\hat{b}^{\dagger} =\frac{1}{\sqrt{2}}( \hat{q}-i\hat{p} ),
\label{anni}%
\end{equation}
such that we can rewrite the Hamiltonian as (we set $\hbar=1$)
\begin{equation}
\hat{H} =\omega_0\left(\hat{b}^{\dagger}\hat{b} +
\frac{1}{2}\right)=\omega_0\left(\hat{n}+ \frac{1}{2}\right).
\label{hamil}%
\end{equation}
Eigenstates for the Hamiltonian (\ref{hamil}) are called Fock or
number states:
\begin{equation}
\hat{H}|n\rangle =\omega_0\left(n+ \frac{1}{2} \right)|n\rangle.
\label{eigen}
\end{equation}
Fock states are orthonormal and form a complete basis, such that
any other state of the harmonic oscillator may be written in terms
of them. In particular coherent states may be written as
\begin{equation}
|\alpha\rangle
=e^{-\frac{|\alpha|^2}{2}}\sum^{\infty}_{n=0}\frac{\alpha^n}{\sqrt{n!}}|n\rangle
=\hat{D}(\alpha)|0\rangle \label{coh}
\end{equation}
where
\begin{equation}
\hat{D}(\alpha)=e^{\alpha \hat{b}^{\dagger}- \alpha^*\hat{b}}.
\label{dis}
\end{equation}
Coherent states are eigenstates of the annihilation operator
\begin{equation}
\hat{b}|\alpha\rangle=\alpha|\alpha\rangle.
\end{equation}
and have the property that the motion of the center of mass of the
wave packet obeys the laws of classical mechanics (see for
instance \cite{Meystre})
\begin{equation}
\langle \alpha |\hat{q}|\alpha\rangle =q_c, \qquad \langle \alpha
|\hat{p}|\alpha\rangle =p_c, \qquad \langle \alpha
|\hat{H}|\alpha\rangle =H_c
\end{equation}
where the index $c$ labels the classical variables.
\section{Time dependent harmonic Hamiltonian}
The time dependent harmonic Hamiltonian reads
\begin{equation}
\hat{H} =\frac{1}{2}\left[\hat{p}^2+\Omega^2(t)\hat{q}^2\right].
\label{td}%
\end{equation}
It is well-known that an invariant may be written for this
interaction that has the form
\begin{equation}
\hat{I}=\frac{1}{2}\left[  \left( \frac{\hat{q}}{\rho}\right)
^{2}+(\rho\hat{p}-\dot{\rho}\hat{q})^{2}\right] \label{inva}
\end{equation}
where $\rho$ obeys the Ermakov equation
\begin{equation}
\ddot{\rho}+\Omega^{2}\rho=\rho^{-3}. \label{erma}
\end{equation}
Furthermore, it is easy to show that $\hat{I}$ may be related to
the Hamiltonian (\ref{unity}) by a unitary transformation of the
form
\begin{equation}
\hat{T}=\rho^{\frac{i}{2}\left(\hat{q}\hat{p}+\hat{p}\hat{q}%
+\frac{2\rho\dot{\rho}}{\rho^2-1}\hat{q}^{2}\right)}
\end{equation}
that can be re-written as (see appendix)
\begin{equation}
\hat{T}=e^{i\frac{\ln(\rho )}{2}(\hat{q}\hat{p}+\hat{p}\hat{q}%
)}e^{-i\frac{\dot{\rho}}{2\rho}\hat{q}^{2}}=
e^{i\frac{\ln\rho}{2}\frac{d\hat{q}^2}{dt}}
e^{-i\frac{\dot{\rho}}{2\rho}\hat{q}^{2}},
\end{equation}
as
\begin{equation}
\hat{H}=\omega_0\hat{T}\hat{I}\hat{T}^{\dagger}.
\label{trai}%
\end{equation}
By using Eq. (\ref{eigen}) we can see that
\begin{equation}
\hat{I}\hat{T}^{\dagger}|n\rangle =\omega_0\left(n+ \frac{1}{2}
\right)\hat{T}^{\dagger}|n\rangle \label{eigenI}
\end{equation}
i.e. states of the form
\begin{equation}
|n\rangle_t =\hat{T}^{\dagger}|n\rangle
\end{equation}
are eigenstates of the so-called Lewis invariant. Lewis
\cite{lewis} wrote his invariant in terms of annihilation and
creation operators
\begin{equation}
\hat{I} =\hat{a}^{\dagger}\hat{a} + \frac{1}{2}=\hat{n}_t+
\frac{1}{2}.
\label{hamilI}%
\end{equation}
with
\begin{equation}
\hat{a}
=\frac{1}{\sqrt{2}}\left[\frac{\hat{q}}{\rho}+i(\rho\hat{p}
-\dot{\rho}\hat{q} )\right], \qquad
\hat{a}^{\dagger}=\frac{1}{\sqrt{2}}\left[\frac{\hat{q}}{\rho}-i(\rho\hat{p}
-\dot{\rho}\hat{q} )\right].
\label{anniI}%
\end{equation}
Once defined creation and annihilation operators, analogous
equations to the harmonic oscillator with constant frequency may
be obtained for the TDHO, for instance
\begin{equation}
\hat{D}_t(\alpha)=e^{\alpha \hat{a}^{\dagger}- \alpha^*\hat{a}}.
\label{disI}
\end{equation}
and
\begin{equation}
\hat{a}|\alpha\rangle_t=\alpha|\alpha\rangle_t.
\end{equation}
with
\begin{equation}
|\alpha\rangle_t
=e^{-\frac{|\alpha|^2}{2}}\sum^{\infty}_{n=0}\frac{\alpha^n}{\sqrt{n!}}|n\rangle_t
=\hat{D}_t(\alpha)|0\rangle_t \label{cohI}.
\end{equation}

Recently it has been shown that the Schr\"odinger equation for the
time dependent harmonic Hamiltonian has a solution of the form
\cite{manu}
\begin{equation}
|\psi(t)\rangle=e^{-i\hat{I}\int_0 ^t\omega(t){dt}}
\hat{T}^{\dagger}\hat{T}(0)|\psi(0)\rangle \label{solut}.
\end{equation}
with $\omega(t)=1/\rho^2$. If we consider the initial state to be
\begin{equation}
|\psi(0)\rangle=
\hat{T}^{\dagger}(0)|\alpha\rangle=|\alpha\rangle_0 \label{ini}
\end{equation}
with $\alpha$ given in (\ref{coh}), we note that the evolved state
has the form
\begin{equation}
|\psi(t)\rangle= \hat{T}^{\dagger}|\alpha
e^{-i\int_0^t\omega(t')dt'} \rangle=|\alpha
e^{-i\int_0^t\omega(t')dt'} \rangle_t, \label{cohevol}
\end{equation}
this is, coherent states keep their form through evolution.
\section{Minimum uncertainty states}
We  define the operators
\begin{equation}
\hat{Q} =\frac{1}{\sqrt{2}}(\hat{a}+\hat{a}^{\dagger}),
\end{equation}%
and
\begin{equation}
\hat{P} =\frac{1}{i\sqrt{2}}(\hat{a}-\hat{a}^{\dagger}).
\end{equation}%
It is easy to see that they are related to $\hat{q}$ and $\hat{p}$
by the transformations
\begin{equation}
\hat{q} = \hat{T}\hat{Q}\hat{T}^{\dagger}
\end{equation}%
and
\begin{equation}
\hat{p} = \hat{T}\hat{P}\hat{T}^{\dagger}
\end{equation}
and that they obey the equations \cite{fer4}
\begin{equation}
\dot{\hat{P}} =i\omega(t)[\hat{I},\hat{P}]=-\omega(t)\hat{Q},
\end{equation}
and
\begin{equation}
\dot{\hat{Q}} =i\omega(t)[\hat{I},\hat{Q}]=\omega(t)\hat{P}.
\end{equation}
 The uncertainty relation for
operators $\hat{Q}$ and $\hat{P}$ for the coherent state
(\ref{cohI}) is given by
\begin{equation}
\Delta\hat{Q}\Delta\hat{P} = \frac{1}{2},
\end{equation}
where $\Delta \hat{X}= \sqrt{\langle \hat{X^2}\rangle-\langle
\hat{X}\rangle^2}$. (Time dependent) Coherent states are thus
minimum uncertainty states (MUS), not for position and momentum
but for the transformed position and momentum.

\section{step function}
Let us consider the Hamiltonian of the system to be given by
equation (\ref{td}) with $\Omega(t)$ given by a step function that
may be modeled by \cite{fer3}
\begin{equation}
\Omega(t) = \omega_1 \left[1+
\frac{\Delta}{2}\left(1+\frac{1}{2}\tanh[\epsilon(t-t_s)]\right)\right]
\end{equation}
where $t_s$ is the time at which the frequency is changed,
$\Delta=\omega_2-\omega_1$ with $\omega_1$ and $\omega_2$ the
initial and final frequencies, and $\epsilon$ is a parameter
($\epsilon \rightarrow \infty$ models better the step function).
In fig. (\ref{fig1}) we plot this function as a function of $t$
for $\omega_1=1$ and $\omega_2=2$ (solid line) and $\omega_2=3$
(dash line). The solution to the Ermakov equation (\ref{erma}) for
this particular form of $\Omega(t)$ is given by \cite{fer3}
\begin{equation}
\rho(t) = \frac{1}{\sqrt{2}}\sqrt{1 +
\frac{\omega_1^2}{\Omega^2(t)}+ \left(1-
\frac{\omega_1^2}{\Omega^2(t)}\right)\cos\left(2\int_{t_s}^t\Omega(t')dt'\right)}.
\end{equation}
A plot of $\rho(t)$ is given in fig. \ref{fig2} for the same
values as Fig. (\ref{fig1}). We also plot $\omega(t)=1/\rho^2$ in
fig. (\ref{fig3}). It may be numerically shown that the time
average of $omega(t)$ from $t_s$ to the end of the first period is
$2$ for the solid line and $3$ for the dash line, with
$\omega_{max}=2^2$ for the solid line and $\omega_{max}=3^2$ for
the dash line.

Let us consider that at time $t=0$ we have the system in the
initial state (\ref{coh}). From \ref{fig2} we can see that
$\hat{T}(0)=1$ as $\dot{\rho}=0$ and $\ln\rho=0$. Therefore from
(\ref{ini})
$|\psi(0)\rangle=\hat{T}^{\dagger}(0)|\alpha\rangle=|\alpha\rangle_0=|\alpha\rangle$
and from (\ref{solut}) we obtain the evolved wave function

\begin{equation}
|\psi(t)\rangle=e^{-i\hat{I}\int_0 ^t\omega(t){dt}}
\hat{T}^{\dagger}|\alpha\rangle = \hat{T}^{\dagger}|\alpha
e^{-i\int_0 ^t\omega(t){dt}}\rangle.
\end{equation}
Note that the coherent state in the above equation is given in the
original Hilbert space, i.e. in terms of number states given in
(\ref{eigen}). From fig. \ref{fig2} we can also see that for the
maximums, $\dot{\rho}(t_{max})=0$ and $\ln\rho(t_{max})=0$,
therefore $\hat{T}^{\dagger}(t_{max})=1$ and
\begin{equation}
|\psi(t_{max})\rangle=|\alpha
e^{-i\int_0^{t_{max}}\omega(t)dt}\rangle,
\end{equation}
i.e. we obtain back the initial coherent state. However, for the
minimums, we have $\dot{\rho}(t_{min})=0$ and
$\ln\rho(t_{min})\neq0$ and then we obtain
\begin{equation}
|\psi(t_{min})\rangle=e^{\frac{i\ln\rho(t_{min})}{2}(\hat{q}\hat{p}+\hat{p}\hat{q})}
|\alpha e^{-i\int_0^{t_{min}}\omega(t)dt}\rangle,
\end{equation}
that may be written in terms of annihilation and creation
operators as
\begin{equation}
|\psi(t_{min})\rangle=e^{\frac{\ln\rho(t_{min})}{2}(\hat{b}^2-(\hat{b}^{\dagger})
^2)}|\alpha e^{-i\int_0^{t_{min}}\omega(t)dt}\rangle,
\end{equation}
that are the well-known squeezed (two-photon coherent) states
\cite{yuen}
\begin{equation}
|\psi(t_{min})\rangle=|\alpha
e^{-i\int_0^{t_{min}}\omega(t)dt};\ln\rho(t_{min})\rangle
\end{equation}
that, as coherent states, squeezed states are also MUS. However
the uncertainties for $\hat{q}$ and $\hat{p}$ are

\begin{equation}
\Delta\hat{q} = \frac{1}{\sqrt{2}\rho(t_{min})},
\end{equation}
and
\begin{equation}
\Delta\hat{p} = \frac{\rho(t_{min})}{\sqrt{2}}.
\end{equation}
For times in between we will have neither coherent states nor
standard squeezed states (in the initial Hilbert space), but the
wave function
\begin{equation}
|\psi(t)\rangle=e^{-i\frac{\dot{\rho}}{2\rho}\hat{q}^2}|\alpha
e^{-i\int_0^{t}\omega(t')dt'}; \ln\rho\rangle.
\end{equation}
It should be stressed however, that in the {\it instantaneous}
Hilbert space we will always have the coherent state
(\ref{cohevol}).

\section{Conclusions}
We have studied the problem of the time dependent harmonic
oscillator for a particular form of time dependency, namely the
step function. We have studied it from an invariant point of view
that has made it possible to obtain analytic solutions. We have
shown that, depending on the space we look at the solutions for an
initial coherent state, the state remains coherent or it may
present squeezing. This squeezing may be enhanced by making larger
the change in frequency.

\section{appendix}

The unitary operator $\hat{T}$ is given by
\begin{equation}
\hat{T}=\rho^{\frac{i}{2}\left(\hat{q}\hat{p}+\hat{p}\hat{q}+
\frac{2\rho\dot{\rho}}{\rho^2-1}\hat{q}^{2}\right)}=e^{i\frac{
\ln\rho}{2}(\hat{q}\hat{p}+\hat{p}\hat{q}+
\frac{2\rho\dot{\rho}}{\rho^2-1}\hat{q}^{2})}.
\end{equation}
The exponential above has the sum of the operators
\begin{equation}
\hat{A}=i\frac{\ln\rho}{2}(\hat{q}\hat{p}+\hat{p}\hat{q})
\end{equation}
and
\begin{equation}
\hat{B}=i \ln\rho \frac{\rho\dot{\rho}}{\rho^2-1}\hat{q}^{2}.
\end{equation}
One can show that
\begin{equation}
[\hat{A},\hat{B}] = 2 \ln\rho\hat{B}.
\end{equation}
Using the fact that the commutator of the two operators is
proportional to one of the operators, we can factor the
exponential in the form (see for instance \cite{moya})
\begin{equation}
e^{i\frac{\ln\rho}{2}(\hat{q}\hat{p}+\hat{p}\hat{q}+
\frac{2\rho\dot{\rho}}{1-\rho^2}\hat{q}^{2})} = e^{i\frac{\ln\rho}{2}(\hat{q}\hat{p}+\hat{p}\hat{q}%
)}e^{-i\frac{\dot{\rho}}{2\rho}\hat{q}^{2}}
\end{equation}

\begin{figure}[hbt]
\caption{\label{fig1} $\Omega(t)$ as a function of $t$ for
$\omega_1=1$ and $\omega_2=2$ (solid line) and $\omega_2=3$ (dash
line). $\epsilon=20$ and $t_s=2$.}
\end{figure}

\begin{figure}[hbt]
\caption{\label{fig2} $\rho(t)$ as a function of $t$ for
$\omega_1=1$ and $\omega_2=2$ (solid line) and $\omega_2=3$ (dash
line). $\epsilon=20$ and $t_s=2$.}
\end{figure}

\begin{figure}[hbt]
\caption{\label{fig3} $\omega(t)$ as a function of $t$ for
$\omega_1=1$ and $\omega_2=2$ (solid line) and $\omega_2=3$ (dash
line). $\epsilon=20$ and $t_s=2$.}
\end{figure}

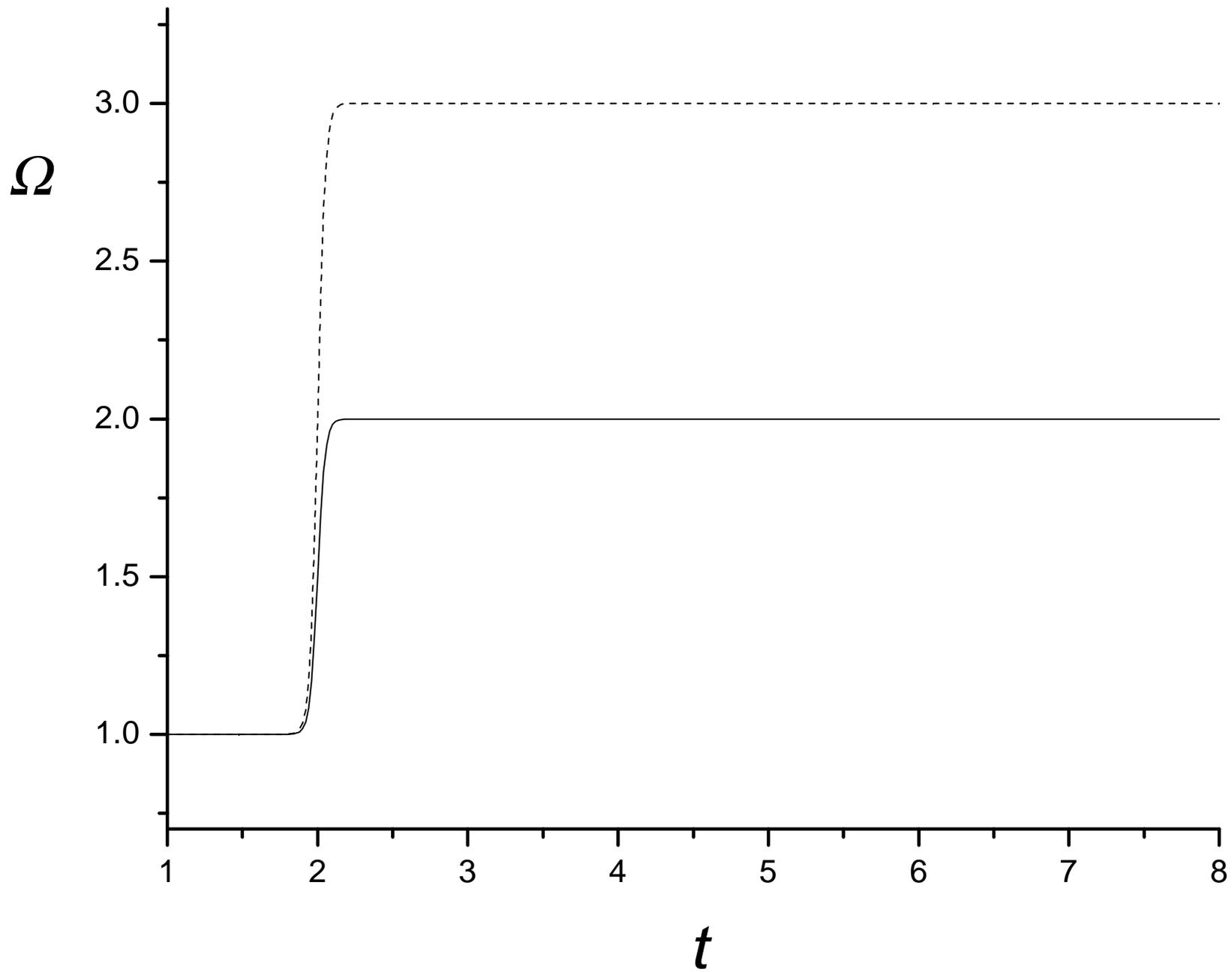
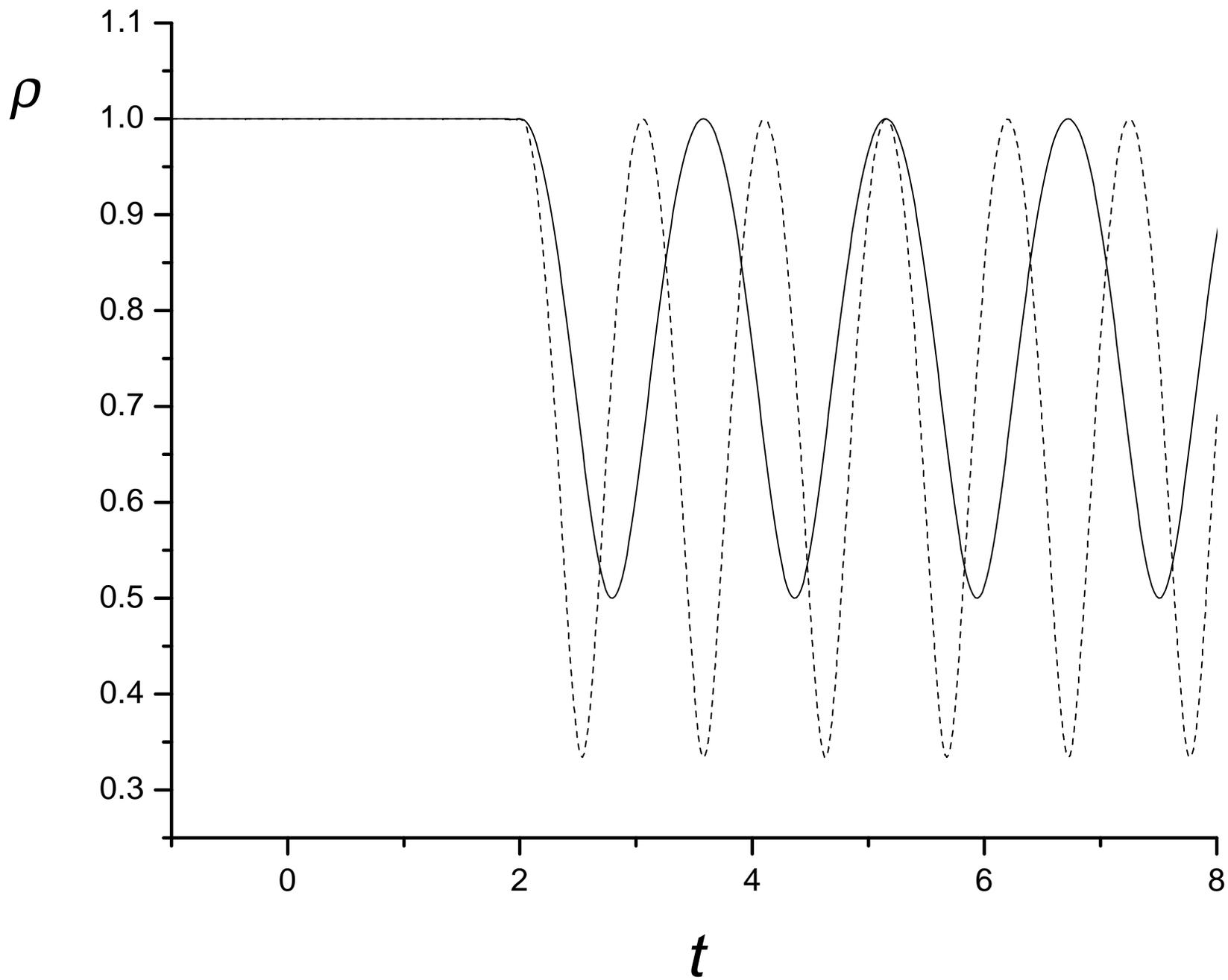
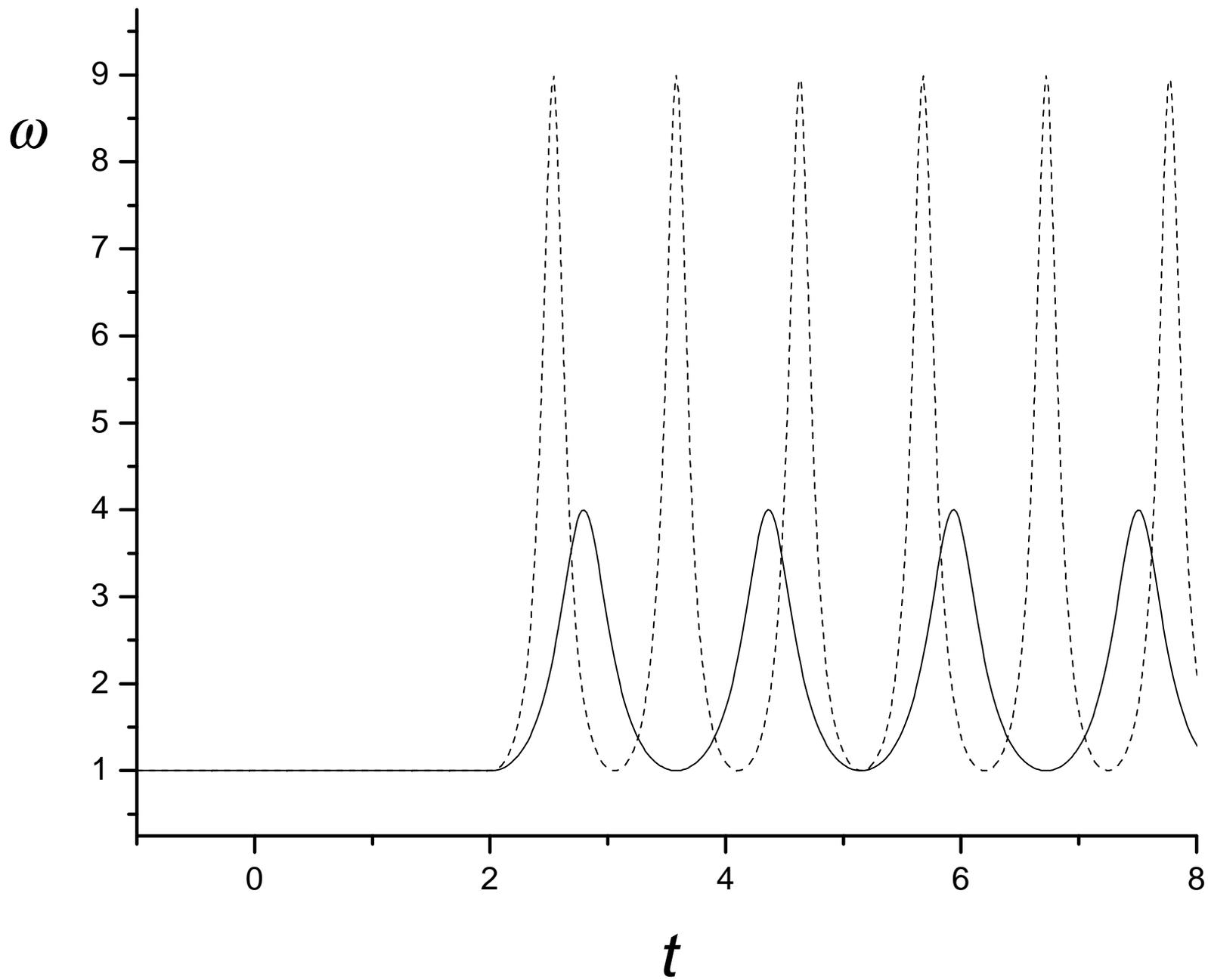
\end{document}